\def \AU{~\rm{AU}}

\documentclass[12pt,preprint]{aastex}
\usepackage{natbib}
\shorttitle{Planets in globular clusters}
\shortauthors{Soker \& Hershenhorn}

\begin{document}

\title{EXPECTED PLANETS IN GLOBULAR CLUSTERS}

\author{Noam Soker\altaffilmark{1} \& Alon Hershenhorn \altaffilmark{1}}
\altaffiltext{1}{Department of Physics, Technion$-$Israel
Institute of Technology, Haifa 32000 Israel;
soker@physics.technion.ac.il}

\begin{abstract}
We argue that all transient searches for planets in globular clusters have
a very low detection probability.
Planets of low metallicity stars typically do not reside at
small orbital separations.
The dependance of planetary system properties on metallicity is
clearly seen when the quantity $I_e \equiv M_p[a(1-e)]^2$ is considered;
$M_p$, $a$, $e$, are the planet mass, semi-major axis, and eccentricity, respectively.
In high metallicity systems there is a concentration of systems at high and low
values of $I_e$, with a low-populated gap near $I_e \sim 0.3 M_J \AU^2$,
where $M_J$ is Jupiter's mass.
In low metallicity systems the concentration is only at the higher range of $I_e$,
{{{ with a tail to low values of $I_e$.}}}
Therefore, it is still possible that planets exist around main sequence stars in
globular clusters, although at small numbers because of the low metallicity,
and at orbital periods of $\ga 10~$days.
We discuss the implications of our conclusions on the role that companions can
play in the evolution of their parent stars in globular clusters, e.g.,
influencing the distribution of horizontal branch stars on the Hertzsprung-Russell
diagram of some globular clusters, {{{ and in forming low mass white dwarfs.}}}
\end{abstract}


\section{INTRODUCTION}
\label{intro}

Evolved sun-like stars that burn helium in their cores occupy the
horizontal branch (HB) in the Hertzsprung-Russel (HR) diagram.
HB stars that have low metallicity and/or low envelope mass are blue, and are
termed blue HB (BHB) stars in globular clusters (GCs), and sdB or sdO
(sdOB together; termed also extreme HB stars or EHB) in the field (not in GCs).
There are strong indications that many of the sdOB stars in the
field are in close binary systems (e.g., Maxted et al. 2001; Maxted 2004),
and there is a strong support to the idea that the sdOB phenomena is
caused by binary companions (e.g., Han et al. 2003; Maxted 2004).
The large fraction of stars in the field that are likely to have planets around
them (Lineweaver \& Grether 2003) hint that planets can also lead to the formation
of sdOB stars (Soker 1998).
We note the recent tentative claim for a planet orbiting an sdB star at an
orbital separation of $a=1.7 \AU$ (Silvotti et al. 2007), that might hint that
closer planets existed in the system before the progenitor turned into a red giant star.

The distributions of stars on the HB (the HB morphology; also
referred to as the color-magnitude diagram [CMD])
differ substantially from one GC to another.
It has long been known that metallicity is the main, but not sole,
factor which determines the HB morphology
(for an historical review see, e.g., Fusi Pecci \& Bellazzini 1997).
The other factor (or factors) which determines the HB morphology is
termed the `second parameter'.
There is a debate on what are the main processes influencing the second parameter
(Catelan 2007 and references therein), with a binary interaction being one of these
processes.
In the low mass binary second parameter model the companion is very light
(a very low mass main sequence [MS] star, a brown dwarf, or a massive planet),
and in most cases will be destroyed as it falls deeper into the envelope
(Soker 1998; Soker \& Harpaz 2007).
Therefore, the non-detection of companions to HB stars in
GCs (Moni Bidin et al. 2006) is not in contradiction with the model.

Although close brown dwarf companions exist (e.g., Zucker \& Mazeh 2000), and are included
in the low mass binary second parameter models, they are rare (e.g., Mazeh et al. 2003;
Grether \& Lineweaver 2006), and so we concentrate on planets and low mass MS stars.

More problematic to the binary second parameter model might seem to be the null detection
of planets to MS stars in GCs (Weldrake et al. 2007b).
Weldrake et al. (2007b) looked for transits in the GCs $\omega$ Cen and 47 Tuc,
and did not find any close planets, but did find stellar companions (Weldrake et al. 2007a).
The null detection of planets in 47 Tuc (Gilliland et al. 2000)
does not contradict the binary model (Soker \& Hadar 2001).
The reason is that this GC has only few stars on its blue HB.
If there were many planets in this GC, then they would be
swallowed by the red giant branch (RGB) star progenitor of the HB star,
causing high mass loss rate (Livio \& Soker 2002) and the formation of many BHB stars.

The null detection of planets in the GCs $\omega$ Cen (Weldrake et al. 2007b) can
be compatible with the planet binary model for the second parameter if
the planets don't reside in close orbits.
Namely, orbital periods $\ga 10~$days.
For that, Soker \& Harpaz (2007) predicted-conjectured that planet
companions might exist in GCs, but at orbital separations of $0.3 \AU \la a \la 3 \AU$.
Planetary systems in the upper orbital separation range will be destroyed in GCs,
but those in the lower range can survive (Fregeau et al. 2006).

Soker \& Harpaz (2007) only based their prediction on the requirement of their model, but did not
bring any observations to support this conjecture.
{{{ Our main goal is to further explore the binary model for the second parameter,
and in particular to examine the possible role of planets.
For that, we turn to examine what can be learned from the wealth of data acquired in the
field of exoplants.}}}
We use recent results {{{ from the field of exoplants}}} to support the claim that if planets exist
in GCs, they are at larger orbital separations than planets around stars close to the sun.
Unlike planets, low mass MS companions can reside close to the parent star
in GCs (Weldrake et al. 2007a), and be much more significant in influencing the
evolution of their parent star.
Namely, the low mass binary second parameter model in low metallicity GCs must be based
mainly on low mass main sequence stellar companions,
but with contribution of massive planets as well.

\section{PLANET COMPANIONS}
\label{planets}

\begin{figure}  %
\vskip 0.1 cm
\resizebox{0.9\textwidth}{!}{\includegraphics{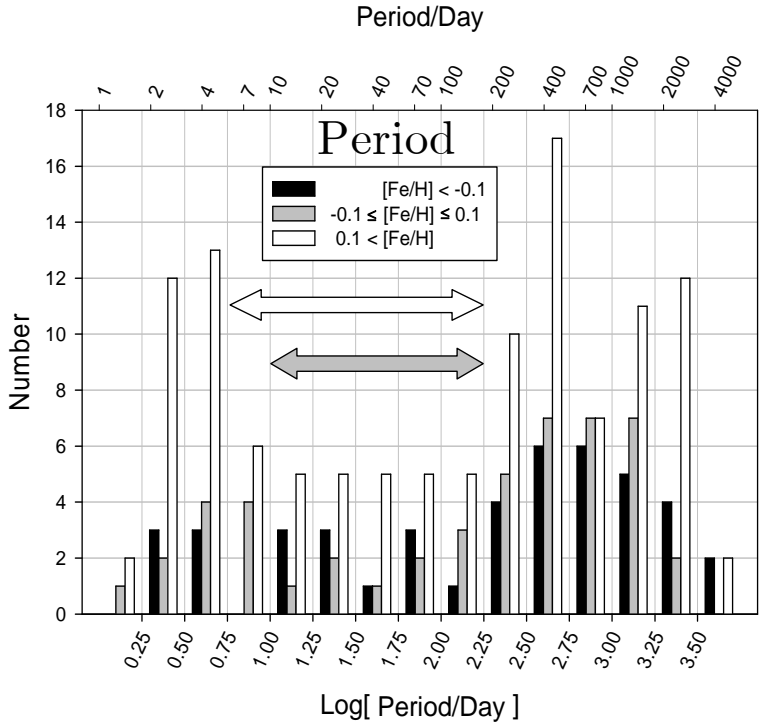}}
\vskip -7.0 cm  
\caption{Histogram of the number of planets as a function of the orbital period $P$
in days, for three ranges of metallicity as indicated.
Each bin shows the number of planets with a period greater than the number to
the left of the bin and smaller than the number to the right of the bin.
The leftmost bin shows the number of planets with a period smaller than the number
to the right of the bin.
The rightmost bin shows the the number of planets with a period greater than the
number to the left of the bin.
{{{ The horizontal thick arrows mark the gaps in the respective two high metallicity ranges.}}}
Data from the Extrasolar Planets Encyclopaedia maintained by Schneider (2007, June 1,
and references therein). }
\label{hisp}
\end{figure}

To support the conjecture that if planets exist in GCs they don't reside at small
orbital separations we present several correlations between properties of known
extra solar planets.
We start by presenting the well known distribution of planets by their orbital period $P$
(Figure \ref{hisp}), but motivated by the recent results of Grether \& Lineweaver (2007)
we do so separately for three ranges of metallicity of the
parent stars: $ {\rm [Fe/H]} < -0.1 $ (black), $-0.1 \le {\rm [Fe/H]} \le 0.1$ (gray),
and $0.1 < {\rm [Fe/H]}$ (white).
Here and in the rest of the diagrams in the paper,
each bin shows the number of planets with a period (or other relevant quantities) greater
than the number to the left of the bin and smaller than the number to the right of the bin.
The leftmost bin shows the number of planets with a period smaller than the number to the right of the bin.
The rightmost bin shows the number of planets with a period greater than the number
to the left of the bin.
All planets data used here are from the Extrasolar Planets Encyclopaedia
maintained by Schneider (2007, and references therein), {{{ as of June 1, 2007.}}}
We skip the comparison of the planet hosting star metallicity distribution with that
of other field stars (see, e.g., Santos et al. 2001).

There is a population gap in this histogram, i.e., the planets are concentrated in two regions.
{{{ The gap exists only for the high and medium metallicity systems, as marked by the two
thick horizontal arrows in Figure \ref{hisp}.}}}
The long period region is limited from above by selection effects.
This gap is well known, e.g., Udry et al. (2003) noted the shortage of planets
in the range $0.06 \AU \le a \le 0.6 \AU$.

\begin{figure}  %
\vskip 0.1 cm
\resizebox{0.9\textwidth}{!}{\includegraphics{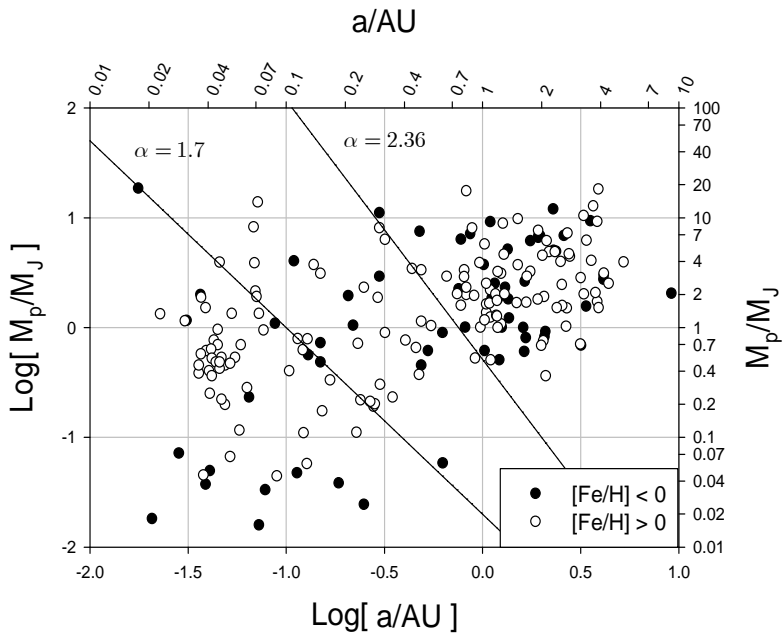}}
\vskip -7.0 cm  
\caption{The distribution of all known planets (from The Extrasolar Planets Encyclopaedia)
in the minimum mass (in $M_J$)$-$semi-major axis (in AU) plane.
Filled and empty circles are for host stars metallicity below and above solar, respectively.
Two lines are drawn to show morphological features in the distribution, with the
value of $\alpha$ marked (eq. \ref{alpha1}).
{{{ The morphological feature we refer to is a low-populated stripe between the
two lines, and two populated clumps, one above the upper line and one below the lower line.}}}  }
\label{mpaf}
\end{figure}

We seek a better quantity to distinguish between close and wide planets,
and between the metallicity ranges.
For that, and motivated by known correlations between planets' mass and period
(e.g., Mazeh et al. 2005), we plot in Figure \ref{mpaf} planets in the $M_p-a$ plane,
where $M_p$ is the minimum planet mass, used here in units of Jupiter mass $M_J$,
and $a$ is the orbital semi-major axis, used here in units of $\AU$.
Filled and empty circles are for systems where the host star metallicity is below
and above solar metallicity, respectively.
We note that there is a morphological structure along a few lines of
\begin{equation}
M_p a^\alpha={\rm constant}.
\label{alpha1}
\end{equation}
Two lines are marked by their $\alpha$ value on Figure \ref{mpaf}.
{{{ These lines bound a low-populated stripe between them, and emphasize two populated clumps:
one consists of planets having high masses and large orbital separations, and the other consists
of planets having low masses and small orbital separations.}}}
Based on this, we take $\alpha=2$ as our standard value to further analyze the correlations.

\begin{figure}  %
\vskip 0.1 cm
\resizebox{0.9\textwidth}{!}{\includegraphics{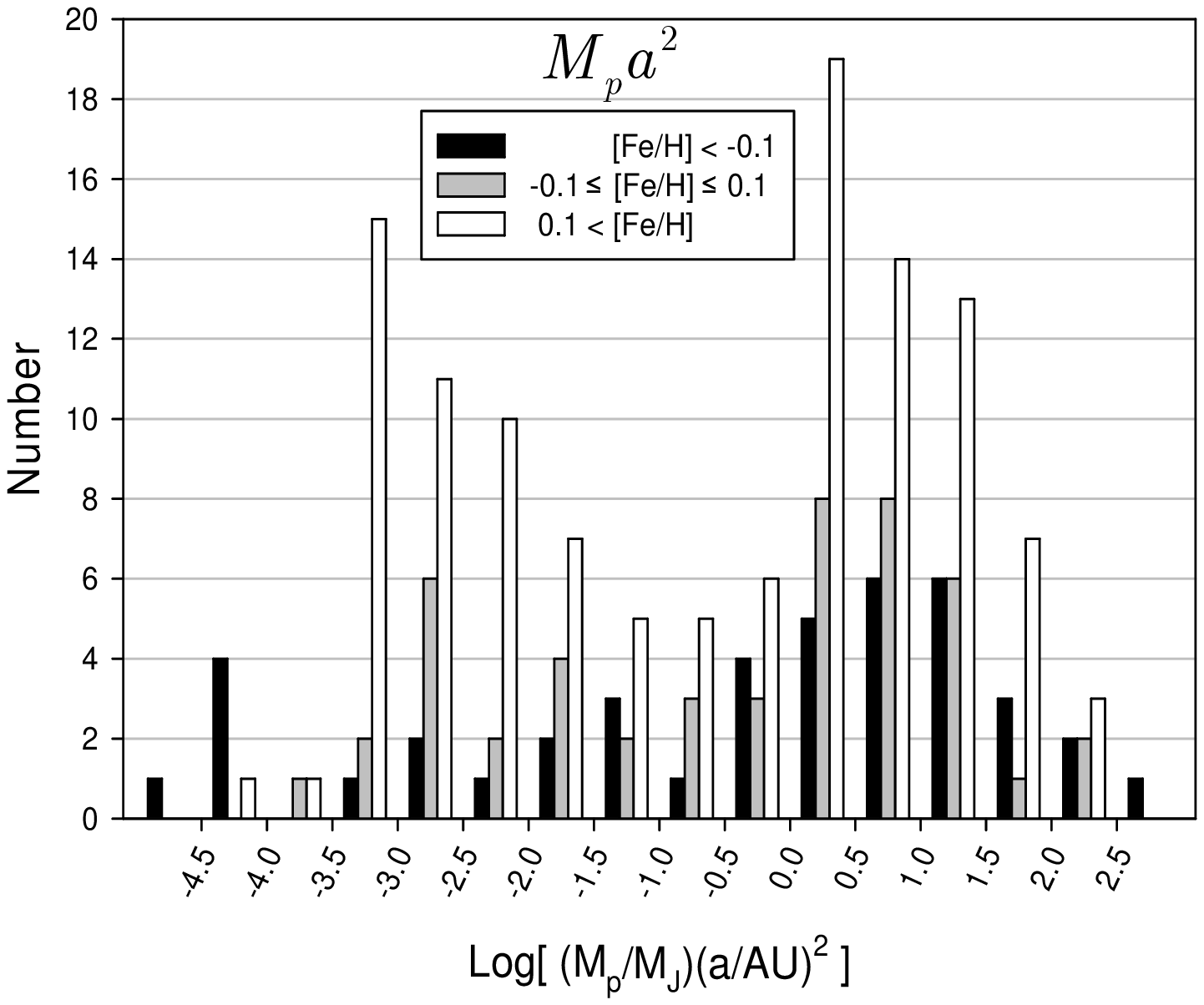}}
\vskip -7.0 cm  
\caption{Histogram of the number of planets as a function
of $M_p a^2$ (in $M_J \AU^2$) for the three metallicity ranges.  }
\label{hismpa}
\end{figure}

\begin{figure}  %
\vskip 0.1 cm
\resizebox{0.9\textwidth}{!}{\includegraphics{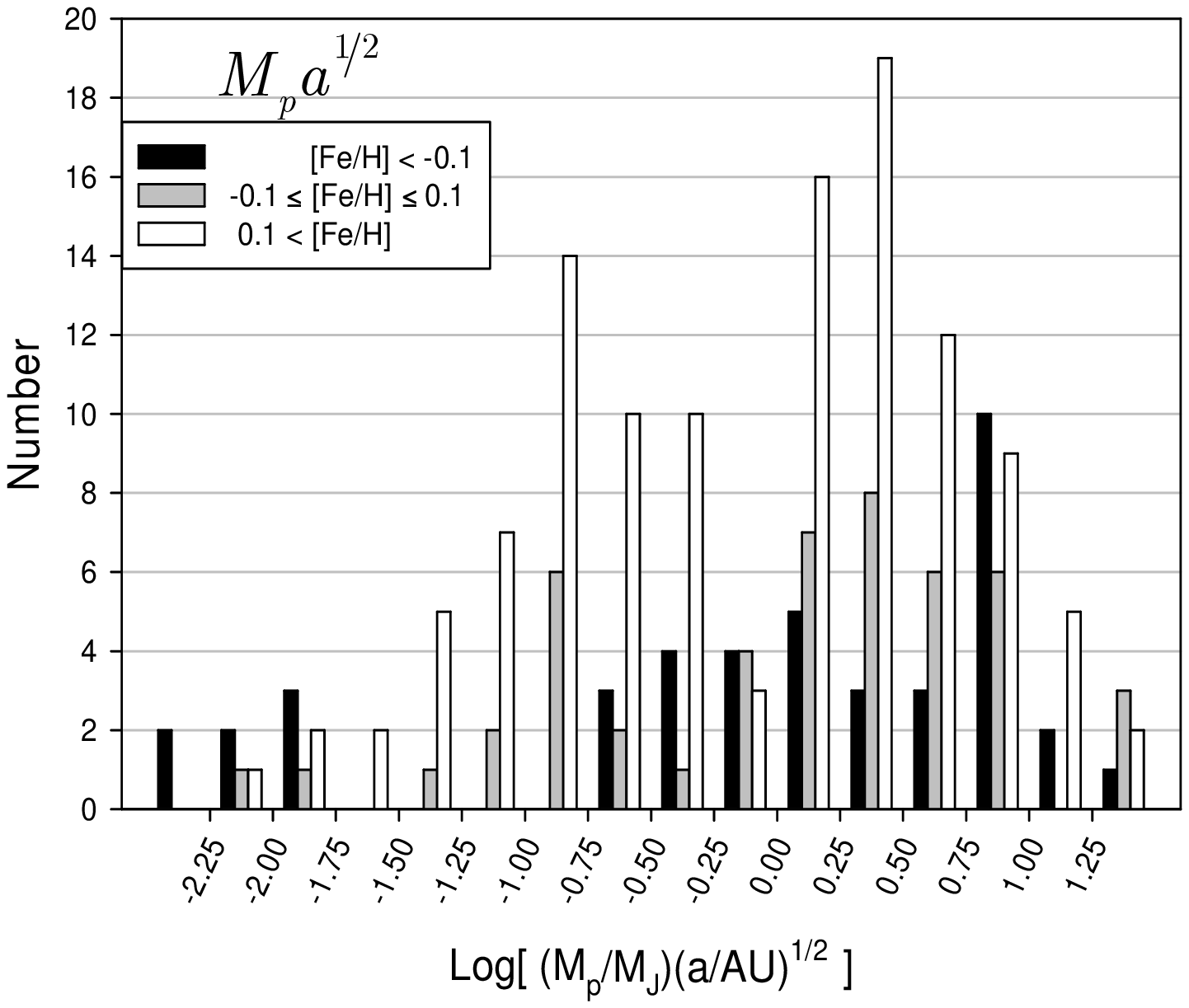}}
\vskip -7.0 cm  
\caption{ {{{ Histogram of the number of planets as a function
of $M_p a^{1/2}$ (in $M_J \AU^{1/2}$) for the three metallicity ranges.}}}  }
\label{hismpraf}
\end{figure}

\begin{figure}  %
\vskip 0.1 cm
\resizebox{0.9\textwidth}{!}{\includegraphics{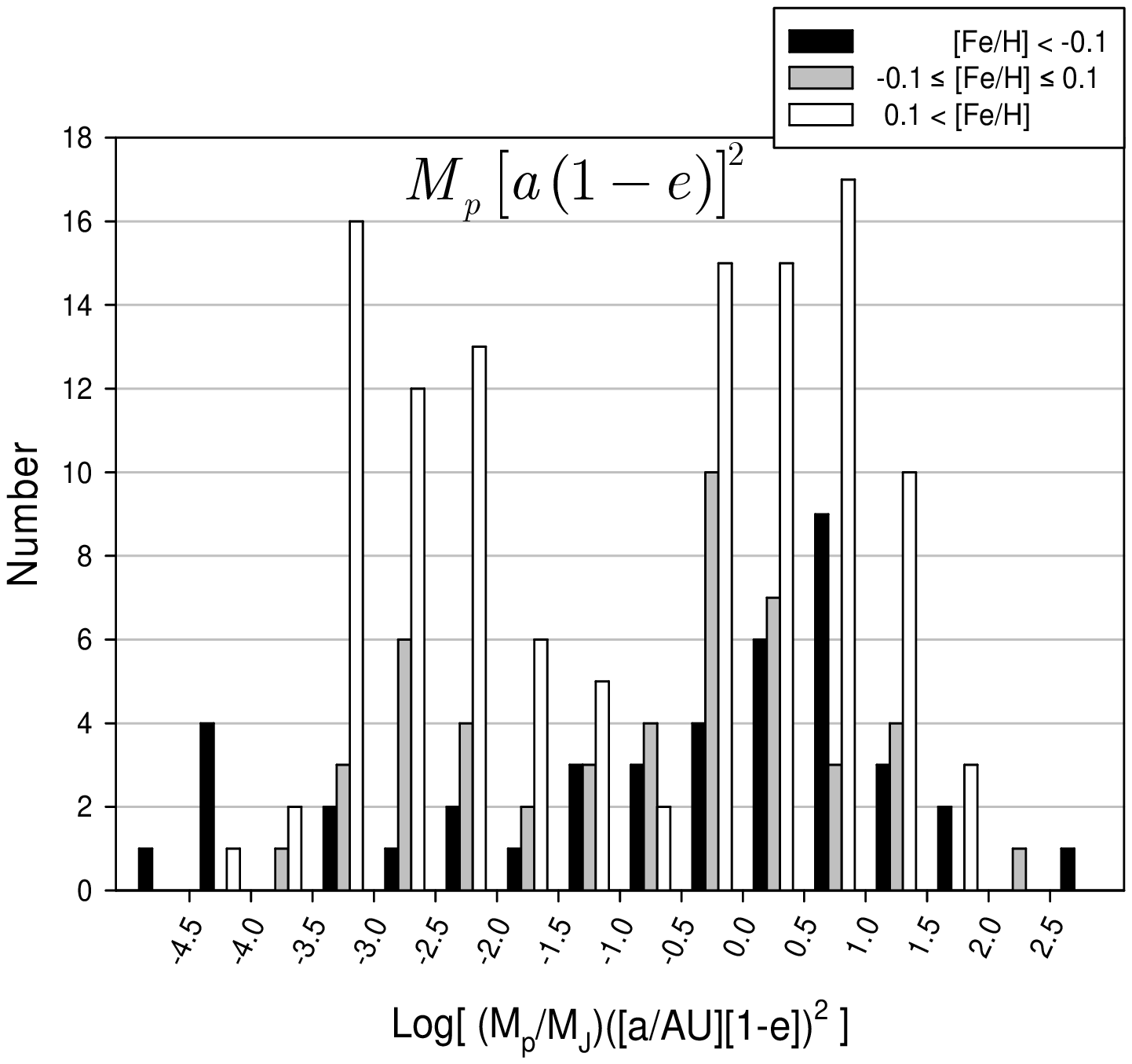}}
\vskip -7.0 cm  
\caption{Histogram of the number of planets as a function
of $I_e=M_p [a(1-e)]^2$ (in $M_J \AU^2$) for the three metallicity ranges.
 }
\label{hismpae}
\end{figure}

\begin{figure}  %
\vskip 0.1 cm
\resizebox{0.9\textwidth}{!}{\includegraphics{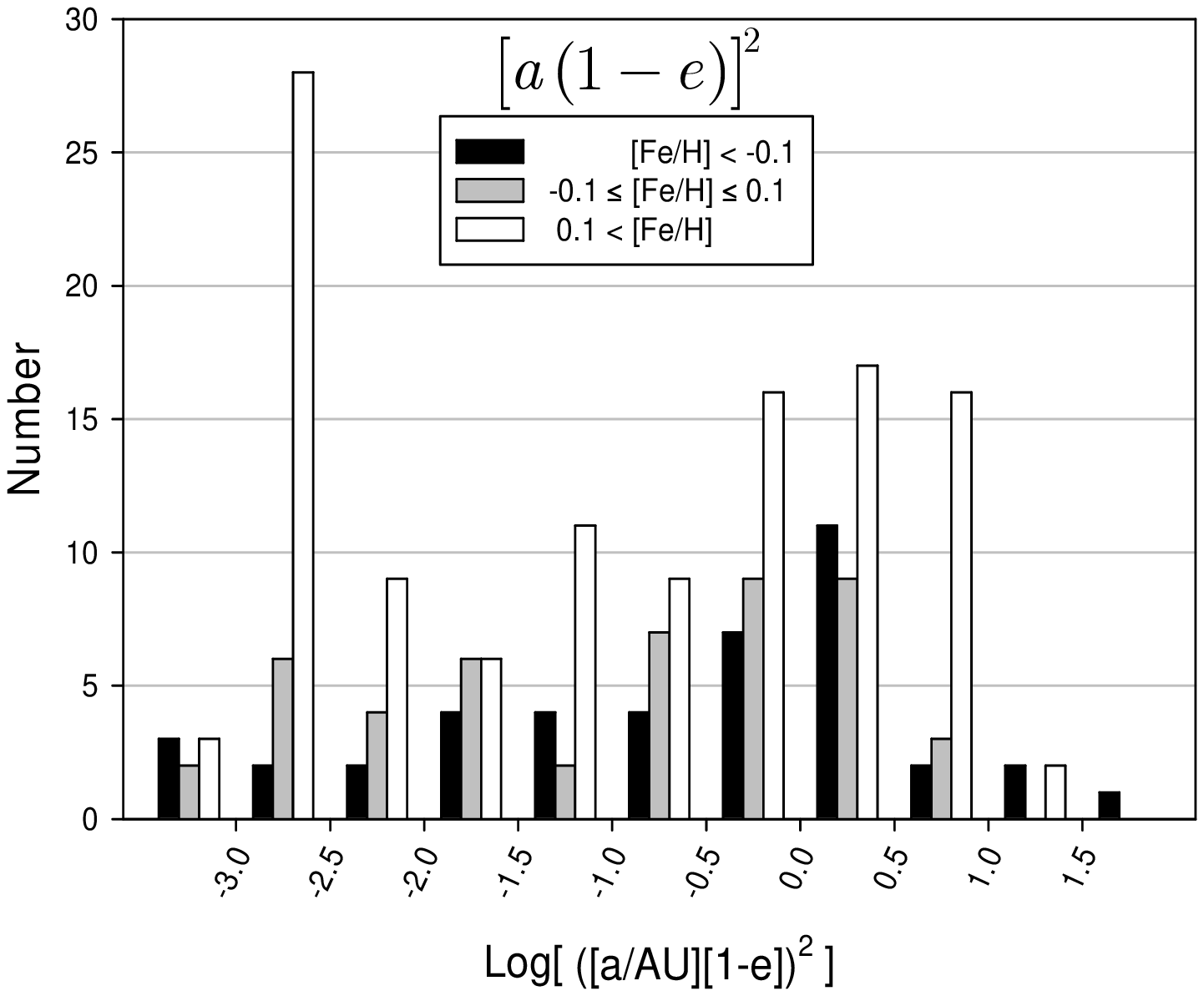}}
\vskip -7.0 cm  
\caption{Histogram of the number of planets as a function
of $[a(1-e)]^2$ (in $\AU^2$) for the three metallicity ranges.
 }
\label{hisae}
\end{figure}

\begin{figure}  %
\vskip 0.1 cm
\resizebox{0.9\textwidth}{!}{\includegraphics{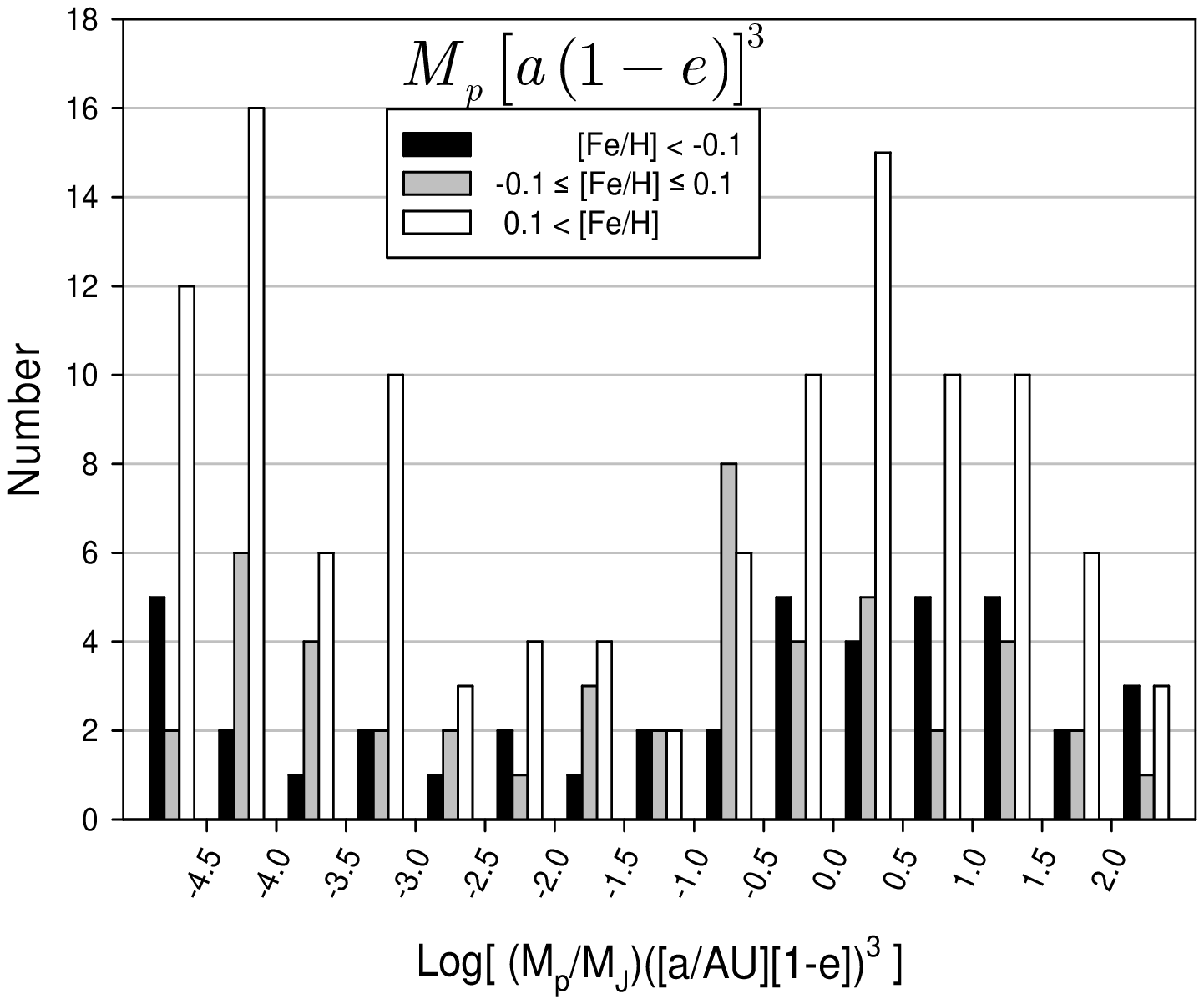}}
\vskip -7.0 cm  
\caption{Histogram of the number of planets as a function
of $M_p [a(1-e)]^3$ (in $M_J \AU^3$) for the three metallicity ranges.
 }
\label{hismpae3}
\end{figure}

\begin{figure}  %
\vskip 0.1 cm
\resizebox{0.9\textwidth}{!}{\includegraphics{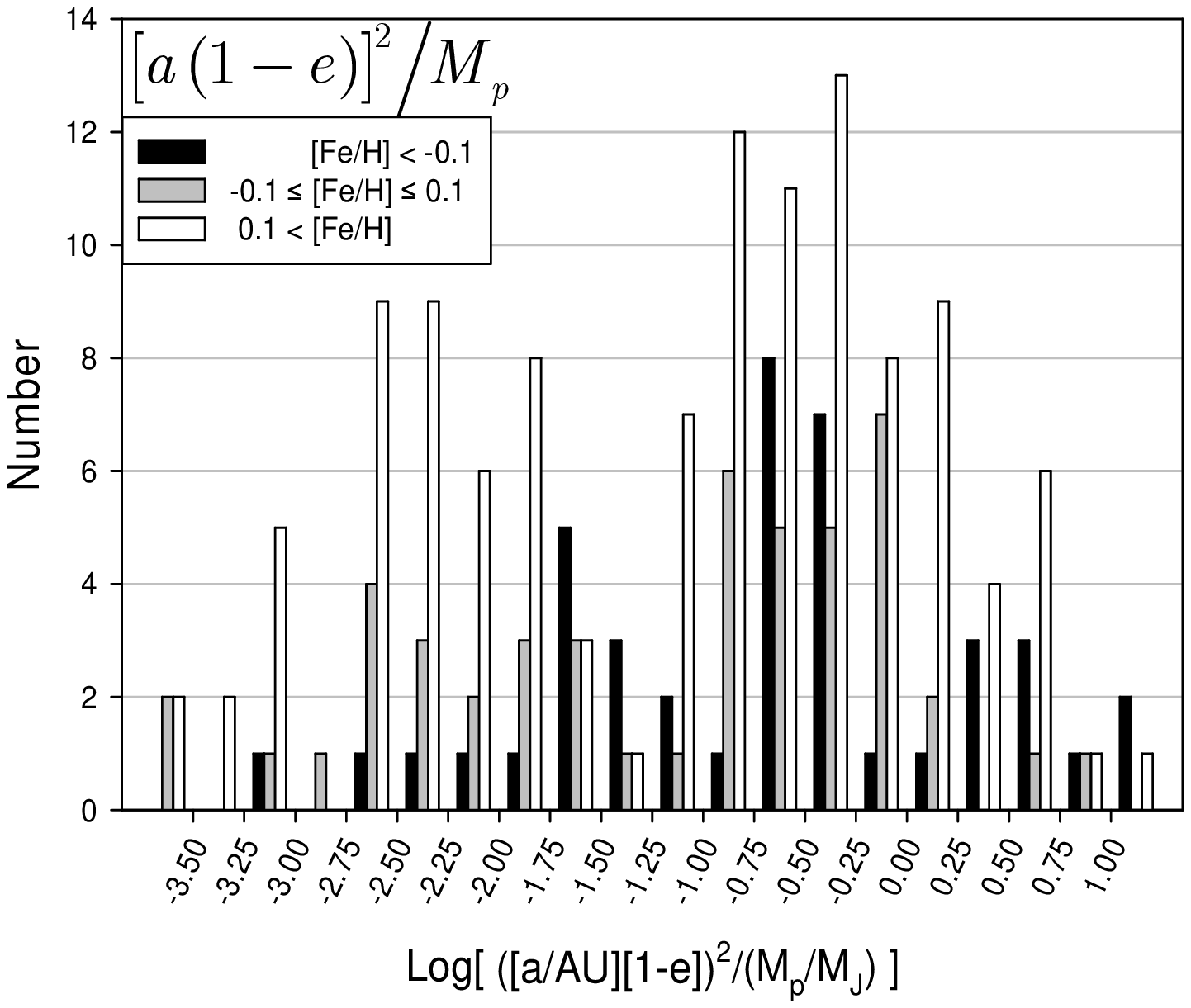}}
\vskip -7.0 cm  
\caption{Histogram of the number of planets as a function
of $[a(1-e)]^2/M_p$ (in $\AU^2~M_J^{-1}$) for the three metallicity ranges.
 }
\label{hismpaed}
\end{figure}

\begin{figure}  %
\vskip 0.1 cm
\resizebox{0.9\textwidth}{!}{\includegraphics{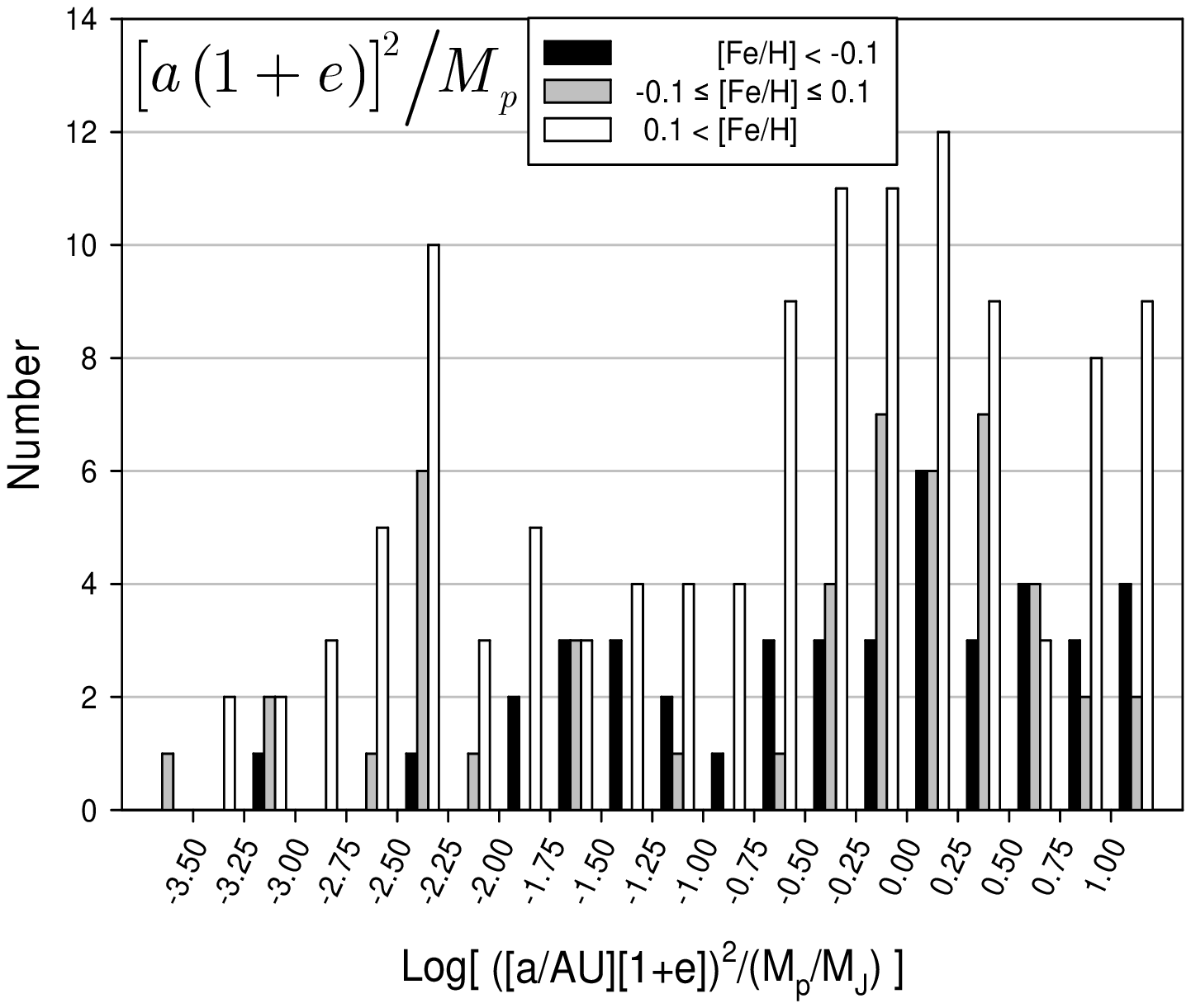}}
\vskip -7.0 cm  
\caption{Histogram of the number of planets as a function
of $[a(1+e)]^2/M_p$ (in $\AU^2~M_J^{-1}$) for the three metallicity ranges.
 }
\label{hismpaepd}
\end{figure}

In Figures \ref{hismpa}-\ref{hismpaepd} we show the distribution
of the entire sample as a function of the following quantities:
$M_p a^2$, $M_p a^{1/2}$,
\begin{equation}
I_e \equiv M_p [a(1-e)]^2,
\label{iee}
\end{equation}
$[a(1-e)]^2$, $M_p [a(1-e)]^3$, $[a(1-e)]^2/M_p$, and $[a(1+e)]^2/M_p$,
respectively. Planets with unknown eccentricity were calculated with $e=0$.
{{{ The quantity $I_e$ has the dimension of moment of inertia, and might therefore
indicate the importance of some kind of interaction between the planet and the parent star,
as will be discussed in section 4.3. As the strongest interaction occurs near periastron,
the relevant distance is $a(1-e)$ rather than $a$ alone.}}}

From Figures \ref{hisp}-\ref{hismpaepd} we learn the following.
\begin{enumerate}
\item
As is well known (e.g., Udry et al. 2003; Marcy et al. 2005) there is a concentration of planets
at very short periods of days. Then there is a low-populated range, the gap, and a
rise to a second grouping of planets at hundreds to thousands of days.
{{{ The gap exists only at the two higher metallicity ranges,
as marked on Figure  \ref{hisp} by the horizontal thick arrows.}}}
\item At lower metallicities planets tend to have longer orbital periods.
The ratio of planets with $P>100~$day to planets with $P<100~$day is
$28/16 = 1.75$ and $31/17 = 1.8$ for $ {\rm [Fe/H]} < -0.1 $ and $-0.1 \le {\rm [Fe/H]} \le 0.1$,
respectively, while it is only $64/53 = 1.2$ for $0.1 < {\rm [Fe/H]}$.
{{{ This trend was found also by Marchi (2007) in the $C2$ and $C3$ sub samples defined there.
This trend will have to be checked with much larger samples in the future.}}}
\item We find that the double-peak distribution of planets at high metallicity is more
pronounced when instead of the period other quantities are used, e.g.,
$ M_p a^2$ or $[a(1-e)]^2/M_p$, or $[a(1+e)]^2/M_p$,
and even more so when the quantity $I_e = M_p [a(1-e)]^2$ is used.
\item From the different parameters we have tried, $I_e= M_p [a(1-e)]^2$ shows most
clearly the two groups of planets at high metallicity, and the differences between planets
with different metallicity of their parent star.
{{{ The criteria we use to prefer the parameter $I_e$ are
(a) a smooth variation in the peaks of the histogram, i.e., small fluctuations in peaks
(e.g., the peaks of the graph of $[a(1-e)]^2/M_p$ in Fig. \ref{hismpaed} have larger fluctuations
than the peaks of $I_e$);
(b) a sharp jump between the group of planets with large separations and the deep gap for
the two high metallicity ranges
(e.g., for $M_p a^2$, presented in Fig. \ref{hismpa}, the difference between the peaks
and gap is not as sharp as for $I_e$ presented in Fig. \ref{hismpae});
(c) A wide gap (e.g., for $M_p a^{1/2}$ presented in Fig. \ref{hismpraf} the gap is very narrow);
and (d) A clear different behavior of the
low metallicity range and the two high metallicity ranges. In particular,
when using $I_e$ such a jump does not exist for the low metallicity range.}}}
For the highest metallicity range used here there is a large jump at
$I_e \simeq 0.3 M_J \AU^2$ ($\log I_e=-0.5$), which clearly separates two $I_e$-populations of planets.
While in high metallicity systems there are two well defined
populations of planets, in low metallicity systems there is only one peak:
Planets of low metallicity stars have typically larger orbital separations.
Because $M_p$ is the minimum mass, in the histogram showing $[a(1-e)]^2/M_p$
(Fig. \ref{hismpaed}) of the real distribution, systems will move to the left
smearing the peak of planets in the lowest metallicity range.
In the histogram using $I_e$ (Fig. \ref{hismpae}), on the other hand,
using the real mass will move systems to the right. This might make
the peak on the right for the lowest metallicity range more pronounced.
\item We have tried to use the same quantities listed above with $1+e$ instead of $1-e$.
For most cases the separation between close and wide planets is worse than
when $1-e$ is used. This implies that the periastron is physically more influential
than the apastron for defining close and wide planets.
However, for the quantity $[a(1+e)]^2/M_p$ shown in figure \ref{hismpaepd} a partition
to two groups is evident. Still, we regard $I_e$ to be the best indicator of the
two planet populations.
\end{enumerate}

{{{
Burkert \& Ida (2007) find that a gap in the semimajor axis distribution does
not exist if only a sample of systems with hosting stellar mass
of $M<1.2 M_\odot$ is used.
We find that when using only hosting stars with masses of $M<1.2 M_\odot$ a gap
does exist in the $I_e$ distribution for the high metallicity range.
There is a gap in the medium metallicity range as well.
We also find that $\sim 70 \%$ of the hosting stars with masses of $M>1.2 M_\odot$
are in our high metallicity range.
The planets in lower metallicity systems with stellar mass of $M>1.2 M_\odot$
tend to have larger values of $I_e$ than planets in systems with stellar
mass of $M  < 1.2 M_\odot$.
We therefore propose that both stellar mass (Burkert \& Ida 2007) and metallicity are
fundamental quantities influencing the distribution of planets around stars.

Burkert \& Ida (2007) also find that the gap in the high hosting stellar mass sample exists
for planets with $M_p \sin i > 0.8 M_\odot$, but not for $M \sin i \le 0.8 M_\odot$.
The dependance on the planet mass is included together with the semimajor axis
and eccentricity in $I_e$.
We find that the $I_e$ distribution does not have a gap for systems with both $M_p >0.8 M_J$ and
$M>1.2 M_\odot$, unlike the semimajor axis distribution found by Burkert \& Ida (2007).
We do see the gap in the $I_e$ distribution of systems with $M_p >0.8 M_J$ and in the $I_e$
distribution of systems with $M < 1.2 M_\odot$.}}}

It is not clear if stars in GCs have planets at all,
{{{ and in particular close-in planets that could be found with the transit technique,}}}
as the fraction
of {{{ detected}}} planetary systems decreases sharply with decreasing metallicity
(e.g., Santos et al. 2001; Fischer \& Valenti 2005; Grether \&  Lineweaver 2007).
{{{ According to Grether \&  Lineweaver (2007) the most probable value of this fraction
for the metallicity range appropriate for GCs is $\la 0.1 \%$, although the uncertainty is large,
and values of $\sim 1 \%$ are still possible.}}}
Contrary to the general trend is the low metallicity of M-dwarfs (low mass MS stars)
hosting planets (Bean et al. 2006), which still leaves hope for planetary systems in GCs.
If some planets do exist in GCs, the implications of the results presented here
are clear: the planets will not be on short orbital periods. Therefore, we conclude,
{\it all transient searches for planets in GCs have a very low detection probability. }
(At least in the statistical sense, as rare close planets might exist.)

{{{ However, planets might be detected in metal-rich clusters.
The open cluster NGC 6791 has [Fe/H]$\sim 0.4$, and has a large population of EHB stars and
low mass WDs (Kalirai et al. 2007), both of which were formed by stars having
increased mass loss on the RGB.
Stellar companions are not likely to cause this increased mass loss (Kalirai et la. 2007).
We propose that the increased mass loss on the RGB in this cluster is partially caused
by planets swallowed by the RGB progenitors of EHB stars and low mass WDs.
We therefore predict that many transient events can be detected in this cluster.}}}

\section{LOW MASS MAIN SEQUENCE COMPANIONS}
\label{mains}

\begin{figure}  %
\vskip 0.1 cm
\resizebox{0.9\textwidth}{!}{\includegraphics{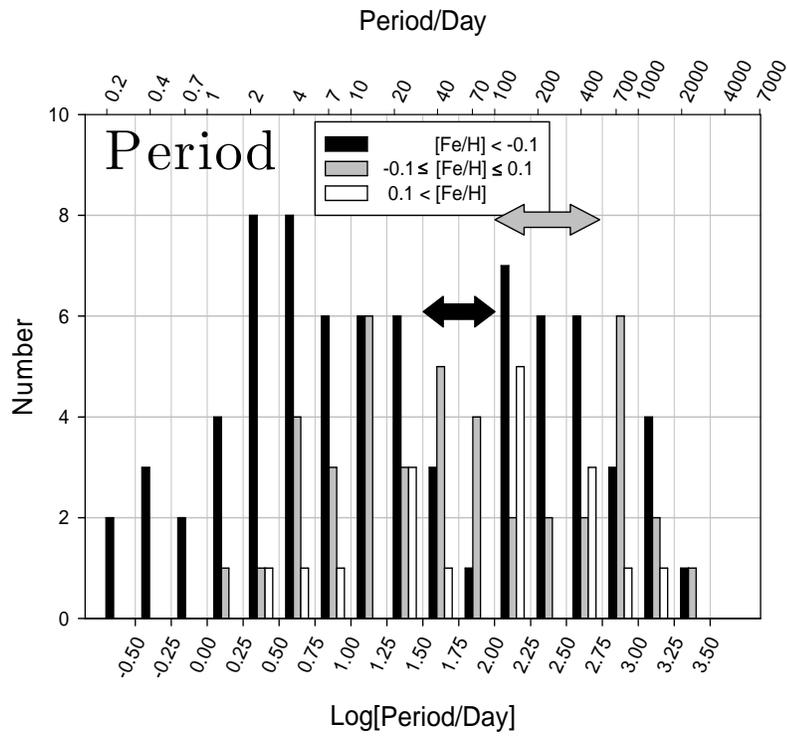}}
\vskip -7.0 cm  
\caption{Histogram of the number of stellar companions as a function of the orbital
period $P$ in days, for three ranges of metallicity as indicated.
Data from Grether \& Lineweaver (2006, 2007). }
\label{pstellar}
\end{figure}

\begin{figure}  %
\vskip 0.1 cm
\resizebox{0.9\textwidth}{!}{\includegraphics{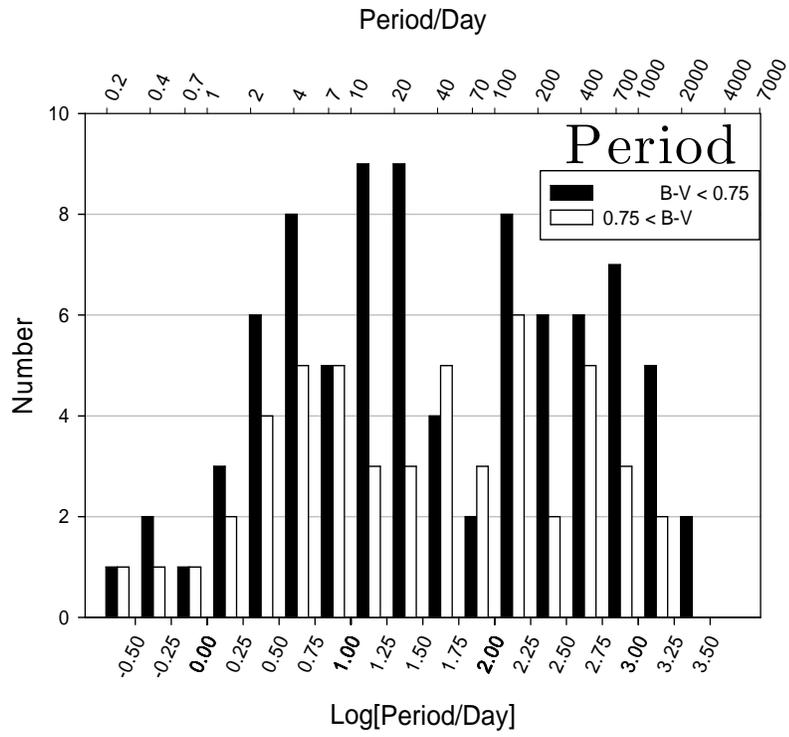}}
\vskip -7.0 cm  
\caption{Histogram of the number of Stellar companions as a function of the orbital
period $P$ in days, for two ranges of color as indicated.  }
\label{BVstellar}
\end{figure}

A similar analysis was conducted as in the previous section but for stellar
companions based on the same sample of 135 systems analyzed by Grether \& Lineweaver (2006).
We present two histograms of the period of the secondary. In figure \ref{pstellar}
the sample was divided into three ranges of metallicity as in the previous section.
In figure \ref{BVstellar} we follow the division of Grether \& Lineweaver (2006) and divide
the sample into two ranges of color of the parent star as marked on the figure.

From figures \ref{pstellar} and \ref{BVstellar} we can deduce the following:
\begin{enumerate}
\item As with planets, there are two concentrations of stellar companions: at short and long
orbital periods.
\item Contrary to the behavior of planets, this double-distribution is more pronounced
in the low and medium metallicity ranges, with longer orbital periods at higher metallicities.
The ratio of stellar companions with
$P>100$~day to $P<100$~day is $27/49 = 0.55$, $15/27 = 0.56$ and
$10/7 = 1.43$ for ${\rm [Fe/H]} < -0.1$, $-0.1 \le {\rm [Fe/H]} \le 0.1$ and $0.1 < {\rm [Fe/H]}$, respectively.
\item Redder systems tend to have a slightly shorter orbital periods.
The ratio of stellar companions with $P > 100$ to $P < 100$~day is $34/50 =0.68$ and $18/33= 0.55$ for
${\rm B-V} < 0.75$ and ${\rm B-V} > 0.75$, respectively.
\item  The width of the gap between long and short orbital periods is $\sim 70$
(from $\sim 30$ to $\sim 100~$day) and $\sim 460$~day (from $\sim 100$ to $\sim 550~$day)
for ${\rm [Fe/H]} < -0.1$ and $-0.1 \le {\rm [Fe/H]} \le 0.1$, respectively.
The center of the gap is located at $P \approx 56$ and $P \approx 247$~day for
${\rm [Fe/H]} < -0.1$ and $-0.1 \le {\rm [Fe/H]} \le 0.1$, respectively.
\end{enumerate}

We have tried to use a finer classification of metallicity and color ranges,
namely ${\rm [Fe/H]} < -0.3$, $-0.3 < {\rm [Fe/H]} < -0.1$,
$-0.1 \le {\rm [Fe/H]} \le 0.1$, and $0.1 < {\rm [Fe/H]}$, for metallicity,
and ${\rm B-V} < 0.6$, $0.6 < {\rm B-V} < 0.7$, $0.7 < {\rm B-V} < 0.8$, and
$0.8 < {\rm B-V}$ for the color (data not shown).
These did not add new information.
We find that the ratio between the number of systems with $P < 100$ days to
systems with $P > 100$ days is $49/27 = 1.8$, $33/15= 2.2$, and  $28/10=2.8$,
for ${\rm [Fe/H]} < -0.1$, ${\rm [Fe/H]} < -0.3$, and ${\rm [Fe/H]} < -0.4$,
respectively.
This shows that the tendency of low metallicity systems to harbor short
period companions is robust.

Although we find that low metallicity stars tend to be slightly closer (shorter orbital periods)
than higher metallicity systems, this does not automatically imply the same trend to low mass systems.
Maxted  \& Jeffries (2005) find that a large fraction of very low mass stars seem to be in
binary systems, but not very close ones.
In addition, a large fraction of stellar companions to low mass stars can have very low
mass $M_2 <0.3 M_\odot$ (Mazeh et al. 2003), and low mass stellar companions tend to
be at large orbital separation (Grether \& Lineweaver 2007).

\section{DISCUSSION AND SUMMARY}
\label{discusion}

\subsection{Main Results}

As is well known and can be seen in figures \ref{hisp}-\ref{hismpaepd},
there are two regions highly-populated with planets, with a low-populated gap between them.
What we have found here (sections 2 and 3) is the following.
\begin{enumerate}
\item In planets the partition to two groups is more significant for high metallicity systems.
\item From the different quantities we have tried, the quantity $I_e \equiv M_p [a(1-e)]^2$
{{{ both distinguishes between high and low metallicity systems,}}}
and shows best this partition to two planet populations in the high metallicity range.
\item In high metallicity systems planets tend to reside on an average closer orbital periods.
{{{ We note that these two properties depend also on the hosting stellar mass (Burkert \& Ida 2007).}}}
\item This trend for metallicity dependance is opposite for stellar companions (section 3).
\item This trend for stellar companions is mainly due to metallicity and not to the
parent star's mass.
There is only a small difference in the ratio of stellar companions
with $P > 100$ to $P < 100$~day for the two color ranges used.
\end{enumerate}

We note that there are other properties for which stars with planetary systems and with stellar
companions show opposite behavior.
The most important is the finding that as metallicity decreases the star is much more likely
to have a stellar companion than to harbor a planetary system (Grether \& Lineweaver 2007).

Our results have implications for two areas.

\subsection{Implications for Globular Clusters}

Since in galactic GCs the metallicity is very low, {{{ close-in}}} planets are not expected there.
However, if planets do exist around some stars in GCs, they will most likely have large
orbital separations. Therefore, the transit search for planets in GCs
has a very low detection probability, and the non-detection of {{{ close-in}}} planets
should not be considered as evidence against the presence of planets in GCs.
We should stay open to the possibility that wide (large orbital periods) planetary systems
exist in GCs.

The large fraction of low metallicity stars that have stellar companions
(Grether \& Lineweaver 2007) implies that a large fraction of stars in GCs
might have formed with stellar companions around them.
It seems that binary systems are indeed common in GCs (Leigh et al. 2007).

Not only metallicity, but other properties at the formation epoch of globular clusters
(Soker \& Hadar 2001) might determine the presence of companions and planets.
For example, density in star forming regions can determine stellar rotation
(Wolff et al. 2007).

Put together, our results support the low mass binary second parameter model,
but the companions {{{ in low metallicity star clusters}}}
are more likely to be very low mass stellar companions, as
observed by, e.g.,  Weldrake et al. (2007a), rather than massive planets.
In that model a low mass companion (a very low mass main sequence star,
a brown dwarf, or a massive planet) influences the post-main sequence evolution of stars,
in particular the properties of the parent stars on the horizontal branch.
(Soker 1998; Soker \& Harpaz 2007).
{{{ In high metallicity clusters, such as NGC 6791 (Kalirai et al. 2007),
planets might be more important than stellar companions in forming EHB stars and low mass
(undermassive) white dwarfs.
We predict that many transient events can be detected in this cluster.}}}

\subsection{Implications for Planetary Systems}

This topic is beyond our scope.
However, our findings suggest that the migration of planets from large to small
orbital separation depends on a combination of parameters expressed by $I_e$ (eq. \ref{iee}).
This quantity has the dimension of moment of inertia, and may imply that a process
reminiscent of the Darwin instability (e.g., Eggleton 2006) is in operation.
In particular, the strong dependance on eccentricity, in that $1-e$ is
a much better factor than $1+e$ in showing the two planet populations,
suggests that some sort of tidal interaction is operating.

Although the situation cannot be simple, let us try the following.
The left limit of the right populated area in figure \ref{hismpae} is
$I_e \simeq 0.3$.
Let us substitute this in the condition for the Darwin instability to occur
(Eggleton 2006, sec. 4.2)
\begin{equation}
\lambda \equiv \frac {M_1 R_1^2 k^2}{M_p a^2} \frac{\Omega}{\omega}> \lambda_{\rm crit},
\label{dar1}
\end{equation}
where $M_1$ is the stellar mass, $R_1$ is the stellar radius, $k R_1$ the radius of gyration
with $k^2 \simeq 0.05-0.1$ for main sequence stars, $\Omega$ is the stellar angular velocity, and
$\omega=2 \pi /P$ is the mean orbital angular velocity.
The critical value $\lambda_{\rm crit}$ rapidly decreases with increasing eccentricity,
with $\lambda_{\rm crit}=1/3,$ $0.171$, $0.102$, and $0.052$, for $e=0$, $0.3$, $0.4$,
and $0.5$, respectively.
As most planets have $e \la 0.4$, in this range we see that
\begin{equation}
\lambda_{\rm crit} \simeq \frac {1}{3}(1-e)^2 \qquad {\rm for} \qquad e \la 0.5.
\label{dar2}
\end{equation}
Substituting approximation (\ref{dar2}) in equation (\ref{dar1}), and taking
$M_1=1 M_\odot$, $k^2=0.075$, and $\Omega \simeq \omega$,
the condition for the Darwin instability to occur becomes
\begin{equation}
R_1 \ga 8 \left( \frac{I_e}{0.3 M_J \AU^2} \right)^{1/2}
\left( \frac{\omega}{\Omega} \right)^{1/2}
R_\odot
\label{dar3}
\end{equation}
This explanation requires that the planets interact with an inflated pre-main sequence star.
This possibility will be studied in a future paper.

We thank Charles Lineweaver and Daniel Grether for giving us their
data on stellar companions.
We thank David Weldrake for useful comments,
{{{ and an anonymous referee for comments that improved the presentation of our results.}}}
This research was supported by a grant from the Asher Space Research
Institute at the Technion.

\end{document}